\documentclass[JHEP]{article}
\pdfoutput=1 % if your are submitting a pdflatex (i.e. if you have
\usepackage{jcappub} % for details on the use of the package, please
\usepackage[T1]{fontenc} % if needed

\usepackage{mathrsfs}
\usepackage{float}
\title{\boldmath Noncanonical Domain Wall as a Unified Model of Dark Energy and Dark Matter: I. Cosmic Dynamics}

\author[a,b,1]{F. A. M. Mulki,\note{Corresponding author.}}
\author[a,b]{H. Wulandari,}
\author[a,b]{T. Hidayat}

\affiliation[a]{Astronomy Research Group,\\ Faculty of Mathematics and Natural Sciences, Institut Teknologi Bandung,\\Jl. Ganesha No. 10 Bandung, Indonesia}
\affiliation[b]{Bosscha Observatory,\\Faculty of Mathematics and Natural Sciences, Institut Teknologi Bandung,\\Jl. Peneropongan Bintang No. 45 Lembang, Indonesia}

\emailAdd{fargiza@as.itb.ac.id}
\emailAdd{hesti@as.itb.ac.id}
\emailAdd{taufiq@as.itb.ac.id}

\abstract{We propose noncanonical domain walls as a new dark energy model inspired by grand unified theories (GUTs). We investigate the cosmic dynamics and discover that the domain walls act as either dark energy or dark matter at different times, depending on the velocity $v$ in the observer's comoving frame. We find a single stable solution to the dynamics, i.e., only freezing ($v=0$) noncanonical domain walls can enter the phantom zone without having to experience ghost field instability. This means that the solution has an equation of state (EoS) $w_{dw}<-1$ without having to possess negative kinetic energy. These domain walls give rise to a late-time cosmic acceleration starting from $z \approx 0.8$, resulting in $w_{dw}=-1.5$ and $w_{eff}=-1.03$ today. We learn that the EoS of the noncanonical domain walls is independent of the potential form. We also investigate the perturbation dynamics following the model. Our simulations show that compared to $\Lambda$CDM, the amplitude of the dark matter power spectrum in the noncanonical domain wall model is lower, while the CMB power spectrum is shifted slightly to lower $l$ multipoles. The proposed model gives a smaller $\sigma_8$ compared to that of $\Lambda$CDM.}

\begin{document}
\maketitle
\flushbottom

\section{Introduction}
\label{sec:introduction}

Following the discovery of cosmic acceleration from supernovae Ia observations \cite{riess_1998,perlmutter_1999}, many subsequent observations confirm the requirement of dark energy to account for the late-time acceleration. The cosmological constant is the simplest and is considered the standard model of dark energy. However, it suffers from serious problems, i.e., fine-tuning and coincidence problems \cite{weinberg_1989}. Therefore, several alternative models have been proposed, either based on formal physics theories or through phenomenological approaches. A class of model that is widely proposed is the scalar field dark energy model, such as quintessence, e.g., \cite{amendola_2000,caldwell_2000}, and k-essence, e.g., \cite{armendariz-picon_2000}. Most alternative models initially focused on solving the cosmological constant problems, such as by introducing attractor solutions and scaling conditions (see,  e.g., \cite{copeland_2006}), but these approaches are less than optimal. Hence, many authors extended the models to include couplings between dark energy and dark matter, e.g., \cite{amendola_2000,pettorino_2008,mulki_2019}.

Currently, new challenges have emerged in cosmology, namely, the conflicts between the results of local measurements of the $H_0$ and $\sigma_8$ parameters and those of CMB Planck observations \cite{riess_2016,riess_2019,planck_2016}. These tensions indicate possibilities of unknown physics, which mainly relate to our lack of understanding of the nature of dark energy. Since the cosmological tensions have become increasingly significant, most proposed models have been intended to ease these tensions. 
The conferred models of dark energy tend to be more phenomenological (see \cite{di-valentino_2017,mortsell_2018,pandey_2020}). This is mainly due to the difficulties encountered by the proposed scalar field class of dark energy in explaining the tensions without giving rise to new theoretical issues. 
However, to elucidate the classical and fundamental problems of dark energy, we require more formal and well-motivated models that stand upon established physics theories.

A grand unified theory (GUT) is a unification theory of fundamental interactions, excluding gravity. This theory explains physical phenomena at high energy levels, one of which is related to the very early universe (see \cite{ross_1985}). One of a GUT's predictions is topological defects emanating from the phase transitions during cosmic inflation. Topological defects of the early universe come in many types, one of which is solitons. A soliton is a solitary wave that behaves like a particle in such a way that satisfies the following conditions: 1) it must maintain its shape while moving at a constant velocity; 2) whenever a soliton interacts with another soliton, they emerge from the collision unchanged except possibly for a phase shift.

A domain wall is a two or three-dimensional soliton that serves as a boundary between topological domains with different energy levels. Domain walls are assumed to be uniformly distributed in space.  In the 2000s, several authors proposed the idea of expressing dark energy as domain walls by employing canonical Lagrangians \cite{fabris_2003,friedland_2002}. Unfortunately, constraints obtained from various observations cause the models to be dropped in the following years. Nevertheless, research on domain walls remains conducted both in particle physics and cosmology. Volkas, in 2011, modeled domain walls in the frame of brane theory and investigated the cosmic dynamics of the model \cite{volkas_2011}. In 2015, Lazanu, Martins, and Shellard examined the effect of domain walls on cosmic dynamics, both as dark energy and as a solitary component of the universe \cite{lazanu_2015}. In addition, the latter work also simulated the CMB power spectrum following the model. Peyravi et al., in 2016, studied spherical domain walls as dark energy inspired by the symmetron model \cite{peyravi_2016}. As mentioned above, the authors employed a canonical Lagrangian, from which the EoS parameter $w \geq -2/3$ cannot reach approximately $ -1$ today. To date, the dynamical properties of domain walls as dark energy following noncanonical formalisms have not been investigated.
Therefore, in this work, we aim to develop a noncanonical formalism for domain walls as dark energy and subsequently investigate their dynamical properties. We also analyze the stability conditions to determine how promising the noncanonical domain walls are if applied to the context of dark energy. The results hopefully can provide a new paradigm for the physics of dark energy and its connection to the predictions of a GUT of the existence of topological defects in the universe \cite{ross_1985}.

This paper is organized as follows.
In Section 2, we describe the formulation of noncanonical formalisms for domain walls, starting from $(2+1)D$ solitons to, by extension, $(3+1)D$ solitons, i.e., the domain walls. Section 3 discusses the dynamics of a homogeneous universe, covering the properties of noncanonical domain walls (phantom and unification properties), the stability conditions of the domain walls, and cosmic dynamics of the background universe. In Section 4, we present the dynamics of the inhomogeneous universe, which includes the analytical derivation of matter perturbation, analytical calculation of the growth rate of the large-scale structures, numerical calculation of matter and the CMB power spectrum, and numerical calculation of $\sigma_8$. Finally, in Section 5, we conclude and discuss future work. Throughout the following discussion, we set the speed of light $c=1$.

\section{Noncanonical formalism for domain walls}
\label{subsec:Lagrangian and equation of motion (EoM)}

In the topological defects physics, for a soliton $\phi$ with a potential $V(\phi)$, a noncanonical Lagrangian in the form of

\begin{equation}
    \mathcal{L} = F(K)-V(\phi),
    \label{eq:L-fang}
\end{equation}
\noindent
where $F(K)$ is an arbitrary function of $K=\frac{1}{2}g^{\mu\nu}\partial_{\mu}\phi\partial_{\nu}\phi$,
has been implemented in modeling spontaneous symmetry breaking resulting in global topological defects called $k$-defects. Depending on the parameters of the Lagrangian, the properties of $k$-defects can be different from the standard/canonical topological defects \cite{babichev2006}.

In the context of cosmic acceleration, Lagrangian (\ref{eq:L-fang}) has been used to generalize a quintessence model by taking $F(K)=K^n$, so that \cite{unnikrishnan2008}

\begin{equation}
    \mathcal{L}=K^n-V(\phi),
	\label{eq:Ln}
\end{equation}

\noindent
where $n$ is an integer and $n>1$. Note that for $n=1$, the usual canonical Lagrangian is retrieved. Lagrangian (\ref{eq:Ln}) has been used to reconstruct a scalar field dark energy potential with a constant equation of state parameter $w_{de}$ \cite{unnikrishnan2008}.
In this work, we use Lagrangian (\ref{eq:Ln}). However, unlike \cite{unnikrishnan2008}, we attempt to retain the topological defect properties of Lagrangian (\ref{eq:Ln}) and study the solitary solutions of the Lagrangian naturally, without any assumptions concerning the dynamics of the equation of state parameter $w_{de}$.

Let us examine the simplest noncanonical case with $n=2$. Hence, with $K=\frac{1}{2}g^{\mu\nu}\partial_{\mu}\phi\partial_{\nu}\phi$, Lagrangian~(\ref{eq:Ln}) becomes

\begin{equation}
    \mathcal{L}=\left( \frac{1}{2}g^{\mu\nu}\partial_{\mu}\phi\partial_{\nu}\phi \right)^{2}-V(\phi),
	\label{eq:L2}
\end{equation}

\noindent
that can further be written as 
\begin{equation}
\begin{split}
    \mathcal{L} &= \frac{1}{4}g^{\mu\nu}\partial_{\mu}\phi\partial_{\nu}\phi g^{\alpha\beta}\partial_{\alpha}\phi\partial_{\beta}\phi - V(\phi)\\
    &= \frac{1}{4}\partial_{\mu}\phi\partial^{\nu}\phi \partial_{\nu}\phi\partial^{\mu}\phi - V(\phi).
	\label{eq:Lurai}
\end{split}
\end{equation}

\noindent
The equation of motion (EoM) for Lagrangian~(\ref{eq:Lurai}) is

\begin{equation}
    \partial_{\mu}\phi\partial^{\mu}\phi\partial_{\nu}\partial^{\nu}\phi + 2\partial^{\mu}\phi\partial_{\nu}\partial_{\mu}\phi\partial^{\nu}\phi + \partial_{\phi}V = 0.
	\label{eq:EoM-KleinGordon}
\end{equation}

By definition, a soliton is a static solution of the scalar wave equation, with the kinetic-time term of the EoM~(\ref{eq:EoM-KleinGordon}) vanishing, or $\partial_{t}\phi=0$. Hence, $\mu$ and $\nu$ in the EoM~(\ref{eq:EoM-KleinGordon}) stand only for space dimensions. Recall that the energy-momentum tensor for an arbitrary Lagrangian $\mathcal{L}$ is given by

\begin{equation}
    T^{\mu\nu}=\frac{2}{\sqrt{-g}}\frac{\delta\left( \sqrt{-g}\mathcal{L} \right)}{\delta{g}_{\mu\nu}}
    = g^{\mu\nu}\mathcal{L} + 2\frac{\delta\mathcal{L}}{\delta{g}_{\mu\nu}}.
    \label{eq:energi-momentum-L}
\end{equation}

\noindent
Applying Lagrangian~(\ref{eq:Lurai}) to Equation (\ref{eq:energi-momentum-L}), we find the energy-momentum tensor for any spacetime metric

\begin{equation}
    T_{\mu\nu} = g_{\mu\nu}V
    + \partial_{\mu}\phi\partial_{\nu}\phi\partial_{\gamma}\phi\partial^{\gamma}\phi
    - \frac{1}{4}g_{\mu\nu}\partial_{\gamma}\phi\partial^{\gamma}\phi \partial_{\delta}\phi\partial^{\delta}\phi.
    \label{eq:energi-momentum}
\end{equation}
\noindent

Before we turn to domain walls, i.e., $(3+1)D$ solitons, let us first examine the dynamics of solitons that occupy time and one-dimensional space, known as kinks. Suppose that the single spatial dimension occupied by the kinks is $x$. Replacing $\mu$ and $\nu$ with $x$, EoM~(\ref{eq:EoM-KleinGordon}) becomes

\begin{equation}
    \partial_{x}\phi\partial^{x}\phi\partial_{x}\partial^{x}\phi + 2\partial^{x}\phi\partial_{x}\partial_{x}\phi\partial^{x}\phi + \partial_{\phi}V = 0.
	\label{eq:EoM-KleinGordon-kink}
\end{equation}

\noindent
For metric $g^{xx}$, we have

\begin{equation}
    \partial_{x}\phi\partial^{x}\phi\partial_{x}{g}^{xx}\partial_{x}\phi + 2g^{xx}\partial_{x}\phi\partial_{x}\partial_{x}\phi{g}^{xx}\partial_{x}\phi + \partial_{\phi}V = 0.
	\label{eq:EoM-KleinGordon-kink-2}
\end{equation}

\noindent
Up to this point, we have not yet specified the spacetime metric. Let us now consider solitons in the cosmological context and manifest solitons as dark energy in a spacetime that represents a homogeneous and isotropic expanding universe. Suppose the universe is well described by a flat  Friedmann-Lema$\hat{\textrm{i}}$tre-Robertson-Walker (FLRW) spacetime with a metric covariance tensor

\begin{equation}
    g_{\mu\nu} = \textrm{diag} \left( -1,a^2,a^2,a^2 \right),
    \label{eq:FLRW-metric}
\end{equation}

\noindent
so that the EoM for the kinks is

\begin{equation}
    3\left( \partial_{x}\phi \right)^2\partial_x^2\phi + a^{4}\partial_{\phi}V
    = 0,
    \label{eq:EoM-FLRW-kink}
\end{equation}

\noindent
where $a$ is the expansion (scale) factor. This equation shows that although the field is definitively static in time, the observer will perceive the dynamic properties of the kinks due to the expansion of the universe. In other words, the kinks are motionless in the comoving frame but are moving in the observer's frame.

We now recollect the Bogomolnyi method introduced in canonical formalisms to reduce the second-order differential equation of EoM into a simple first-order differential equation. Multiplying Equation~(\ref{eq:EoM-FLRW-kink}) with $\partial_{x}\phi$, we obtain

\begin{equation}
    0=3\left( \partial_{x}\phi \right)^3\partial_x^2\phi + a^{4}\partial_{\phi}V\partial_{x}\phi
    = \partial_{x} \left( \frac{3}{4} \left(\partial_{x}\phi\right)^{4} + a^{4}V \right),
\end{equation}

\noindent
and integrating both sides, we find

\begin{equation}
    \frac{3}{4} \left(\partial_{x}\phi\right)^{4} + a^{4}V = C = \textrm{constant}.
    \label{eq: EoM-Bogomolnyi-C}
\end{equation}

We can determine the constant $C$ from the boundary conditions. One of the properties of solitons is that they are localized in space. This property implies that the potential $V$ must vanish when $x\rightarrow\pm\infty$. Meanwhile, $\partial_{x}\phi=\frac{\partial\phi}{\partial x}$ goes to zero when $x\rightarrow\pm\infty$. Therefore, the value of $C$, which localizes the kinks in space, is $C=0$. Thus, the Bogomolnyi equation for the kinks with Lagrangian~(\ref{eq:L2}) is

\begin{equation}
    \partial_{x}\phi = \pm \left( - \frac{4}{3}V \right)^{\frac{1}{4}}a.
    \label{eq:EoM-Bogomolnyi}
\end{equation}

\noindent
Equation~(\ref{eq:EoM-Bogomolnyi}) shows that the size of localization increases with the expansion of the universe. Equation (\ref{eq:EoM-Bogomolnyi}) is imaginary. However, in the next section, we will show that this equation produces real observable quantities.

In principle, a domain wall is an extension of a kink. If we put Equation (\ref{eq:EoM-Bogomolnyi}) in a higher spatial dimension, we will have a planar structure called a planar domain wall, which for brevity, we will call it a domain wall. The energy density of the wall is concentrated in a two-dimensional plane, whose formulation is similar to the kink case. Even though domain walls are an extension of kinks, they have several characteristics that kinks do not have, namely, (see \cite{vachaspati_2006}): 1) domain walls can be curved (because they are 2D or 3D topological objects), and then deformation can occur; 2) domain walls are invariant under Lorentz transformation in the direction perpendicular to the wall planes because the solution does not depend on $t, y$, and $z$. Therefore, any transformation in these coordinates does not affect the shape of the solution.

\section{Homogeneous universe with noncanonical domain walls}

\subsection{Domain walls as a network of kinks}

Domain walls produced by spontaneous symmetry breaking in the early universe have new and dramatic cosmological consequences. If we assume that there exists a dense network of walls within the cosmic horizon, and the network has not lost much of its energy during the history of the universe, then we can estimate the evolution of its energy density.

The energy-momentum tensor of the scalar field $\phi$ forming the wall network is shown in Equation~(\ref{eq:energi-momentum}). Thus, its average over a large volume $\mathcal{V}$ is

\begin{equation}
\begin{split}
    \left\langle T_{\mu\nu} \right\rangle  = \left\langle g_{\mu\nu}V \right\rangle
    + \left\langle \partial_{\mu}\phi\partial_{\nu}\phi\partial_{\gamma}\phi\partial^{\gamma}\phi \right\rangle
     - \left\langle \frac{1}{4}g_{\mu\nu}\partial_{\gamma}\phi\partial^{\gamma}\phi \partial_{\delta}\phi\partial^{\delta}\phi \right\rangle.
    \label{eq:energi-momentum-average}
\end{split}
\end{equation}

If we assume that the field $\phi$ is spatially isotropic, its derivative with respect to spatial coordinates is also isotropic, so

\begin{equation}
    \left\langle \left(\partial_{x}\phi\right)^2 \right\rangle
    = \left\langle \left(\partial_{y}\phi\right)^2 \right\rangle
    = \left\langle \left(\partial_{z}\phi\right)^2 \right\rangle,
    \label{eq:asumsi-isotropis}
\end{equation}

\noindent
and

\begin{equation}
    \left\langle \partial_{i}\phi\partial_{j}\phi \right\rangle = 0,
\end{equation}

\noindent
for $i\neq{j}$. We can write the relation between the mean square of the three-dimensional velocity and the mean square of the one-dimensional velocities as

\begin{equation}
    \left\langle \phi'^{2} \right\rangle = \frac{1}{3} \left\langle \left(\partial_{x}\phi\right)^2 + \left(\partial_{y}\phi\right)^2 + \left(\partial_{z}\phi\right)^2 \right\rangle,
    \label{eq:asumsi-isotropis-2}
\end{equation}

\noindent
where the prime $\left('\right)$ represents a derivative with respect to $r$ defined by $dr^2=dx^2+dy^2+dz^2$.

Generalizing the Bogomolnyi Equation (\ref{eq:EoM-Bogomolnyi}), we realize that the mean square of the three-dimensional velocity in Equation~(\ref{eq:asumsi-isotropis-2}) is equal to the change in the field in the three-dimensional space, which satisfies Equation~(\ref{eq:EoM-Bogomolnyi}). Therefore,

\begin{equation}
\begin{split}
    \phi'& = \pm \left( - \frac{4}{3}V \right)^{\frac{1}{4}}a\\
    \text{or} \;\;
    %\Longleftrightarrow
    \left\langle \phi'^2 \right\rangle & = \left( - \frac{4}{3} \right)^{\frac{1}{2}}a^2\left\langle V^{\frac{1}{2}} \right\rangle.
    \label{eq:energi-kinetik}
\end{split}
\end{equation}

\noindent
Although we have assumed that the field is spatially isotropic, we still use the notation $\left\langle~\right\rangle$ to emphasize that these are averaged quantities. This means that Equation (\ref{eq:energi-kinetik}) holds for averages of the quantities. For $\nu=\mu$, Equation~(\ref{eq:energi-momentum-average}) reads

\begin{equation}
\begin{split}
    \left\langle T_{\mu\mu} \right\rangle = \left\langle g_{\mu\mu}V \right\rangle
    + \left\langle \partial_{\mu}\phi\partial_{\mu}\phi\partial_{\gamma}\phi\partial^{\gamma}\phi \right\rangle
    -\left\langle \frac{1}{4}g_{\mu\mu}\partial_{\gamma}\phi\partial^{\gamma}\phi \partial_{\delta}\phi\partial^{\delta}\phi \right\rangle.
    \label{eq:energi-momentum-average-mu-mu}
\end{split}
\end{equation}

Furthermore, we shall obtain the pressure and the energy density. We first calculate the average of the energy-momentum tensor for $\mu=x$. Equation~(\ref{eq:energi-momentum-average-mu-mu}), broken into its components and employing Equations~(\ref{eq:asumsi-isotropis-2}) and~(\ref{eq:FLRW-metric}), becomes

\small
\begin{equation}
    \begin{split}
    \left\langle T_{xx} \right\rangle = a^2 \left\langle V \right\rangle + \left\langle \phi'^2 \right\rangle
    \left\langle -\left(\partial_{t}\phi\right)^2 + 3a^{-2}\left\langle \phi'^2 \right\rangle \right\rangle
    + \frac{1}{4}a^2\left\langle\left(\partial_{t}\phi\right)^4 - 6a^{-2}\left(\partial_{t}\phi\right)^2 {\left\langle \phi'^2 \right\rangle} + 9a^{-4} {\left\langle \phi'^2 \right\rangle}^2 \right\rangle.
    \end{split}
    \label{eq:energi-momentum-Txx}
\end{equation}
\normalsize

\noindent
By using Equation~(\ref{eq:energi-kinetik}), we can rewrite Equation~(\ref{eq:energi-momentum-Txx}) as

\small
\begin{equation}
\begin{split}
    \left\langle T_{xx} \right\rangle &= -6a^2\left\langle V \right\rangle
    + \frac{a^2}{4}\left\langle\dot{\phi}^4 \right\rangle
    - i\frac{5}{2} \left(\frac{4}{3}\right)^{\frac{1}{2}}a^2 \left\langle V^{\frac{1}{2}} \right\rangle \left\langle \dot{\phi}^2 \right\rangle.
    \label{eq:energi-momentum-xx}
\end{split}
\end{equation}
\normalsize

\noindent
Similarly, for the $tt$-component, we obtain

\begin{equation}
    \left\langle T_{tt} \right\rangle = -4\left\langle V \right\rangle
    - \frac{3}{4}\left\langle\dot{\phi}^4 \right\rangle
    + i\frac{3}{2} \left(\frac{4}{3}\right)^{\frac{1}{2}} \left\langle V^{\frac{1}{2}} \right\rangle \left\langle \dot{\phi}^2 \right\rangle,
    \label{eq:energi-momentum-tt}
\end{equation}

\noindent
where ''\:\:$\dot{}$\:\:'' represents a derivative with respect to cosmic time $t$.

Thus far, we have not specified any fluid property of the domain walls. If we assume that on the cosmological scale, domain walls behave like a perfect fluid with the energy-momentum tensor $T_{\mu\nu}= \textrm{diag}\left(-\rho,P,P,P\right)$, and taking the isotropic assumption in Equation~(\ref{eq:asumsi-isotropis}) into account , we have $\left\langle T_{xx} \right\rangle =\left\langle T_{yy} \right\rangle =\left\langle T_{zz} \right\rangle = P, \textrm{and } \left\langle T_{tt} \right\rangle = -\rho$. Therefore, the energy density and the pressure of the noncanonical domain walls are

\small
\begin{equation}
    \begin{split}
    \rho_{dw} & = 4\left\langle V \right\rangle
    + \frac{3}{4}\left\langle\dot{\phi}^4 \right\rangle
    - i\frac{3}{2} \left(\frac{4}{3}\right)^{\frac{1}{2}} \left\langle V^{\frac{1}{2}} \right\rangle \left\langle \dot{\phi}^2 \right\rangle;\\
    P_{dw} & = -6a^2\left\langle V \right\rangle
    + \frac{a^2}{4}\left\langle\dot{\phi}^4 \right\rangle
    - i\frac{5}{2} \left(\frac{4}{3}\right)^{\frac{1}{2}}a^2 \left\langle V^{\frac{1}{2}} \right\rangle \left\langle \dot{\phi}^2 \right\rangle.
    \label{eq:rho-dan-P}
    \end{split}
\end{equation}
\normalsize

We see that the pressure and density of the domain walls here depend on the average of $V$ and $\dot{\phi}$.
Most domain wall models have a constant $V$. Hence, the time dependence of the two equations above lies in the $\dot{\phi}$ component, and the pressure grows more negative as the universe expands. The imaginary terms in both equations are consequences of the quadratic term in the Lagrangian. We shall show in the following
discussion that an appropriate transformation will cancel
out the imaginary terms in the two equations.

Note that we still let the time derivative terms appear to accommodate the general properties of the domain walls, which have been assumed to follow those of a perfect fluid. However, since a soliton is by definition static, the presence of $\dot{\phi}$ here is supposed to be due to another physical process. In the following subsection, we shall show that a Lorentz boost between the frame of the domain walls and the frame of the observer can be the cause of the presence of $\dot{\phi}$. 

\subsection{Phantom properties of noncanonical domain walls}
\label{subsec:phantom-properties}

We have assumed that the field is spatially isotropic. Therefore, for analytical purposes, we will focus on the motion of a one-dimensional field.
Lagrangian (\ref{eq:Ln}) is invariant under Lorentz transformation \cite{singh_2016}. Therefore, for any potential, the time-dependent solution of the Bogomolnyi Equation (\ref{eq:EoM-Bogomolnyi}) can be obtained by performing a Lorentz boost to the solution of the equation. Consequently, we assume that the time dependence of Equation~(\ref{eq:rho-dan-P}) is caused by the Lorentz boost. We also mentioned that the spatial coordinate that experiences a Lorentz boost, namely the $x$ coordinate, is perpendicular to the wall plane. Suppose that the observer's frame is $O(t,x,y,z)$ and the frame of the domain walls is $D(t_D,x_D,y_D,z_D)$; then, under Lorentz transformation, the two frames are connected via

\begin{equation}
\begin{split}
    t_D=\gamma\left(t-vx\right), \quad x_D=\gamma\left(x-vt\right), \quad 
    y_D=y, \quad z_D=z.
    \label{eq:transformasi-Lorentz}
\end{split}
\end{equation}

\noindent
The subscript $D$ refers to the frame of the domain walls, and $v$ is the relative velocity between the frames of the domain wall and the observer. Making use of Equation~(\ref{eq:transformasi-Lorentz}), we notice that the kinetic component measured by the observer relates to the proper kinetic component in the frame of the domain wall

\begin{equation}
   \dot{\phi} = \partial_t \phi = \frac{\partial \phi}{\partial t} = \frac{\partial \phi}{\partial x_D}\frac{\partial x_D}{\partial t} = -
    \gamma v\partial_{x_D}\phi.
\end{equation}

\noindent
However, $\partial_x\phi=\gamma \partial_{x_D}\phi$, therefore

\begin{equation}
    \dot{\phi} = - v\partial_x\phi,
    \label{eq:hubungan phi dot dan v}
\end{equation}

\noindent
Using Equation~(\ref{eq:EoM-Bogomolnyi}), Equation~(\ref{eq:hubungan phi dot dan v}) becomes

\[ 
\dot{\phi}= \left\{
\begin{array}{ll}
      -v\left(-\frac{4}{3}V\right)^{\frac{1}{4}}a & \textrm{(for positive solution of eq. \ref{eq:EoM-Bogomolnyi})}, \\
      v\left(-\frac{4}{3}V\right)^{\frac{1}{4}}a & \textrm{(for negative solution of eq. \ref{eq:EoM-Bogomolnyi})}. \\
\end{array} 
\right. 
\]
\noindent
Taking the quadratic and quartic of $\dot{\phi}$, we have

\begin{equation}
\begin{split}
    \dot{\phi}^2 &= v^2 \left(-\frac{4}{3}V\right)^{\frac{1}{2}}a^2,\\
    \dot{\phi}^4 &= -\frac{4}{3}v^4Va^4.
    \label{eq:energi-kinetik-kink-non-kanonik}
\end{split}
\end{equation}

Recall that the velocity $v$ due to the Lorentz boost is the relative velocity between the frames of the domain walls and the observer frame, both of which are inertial frames. In other words, the relative velocity is constant over time (hereafter referred to as the velocity of the domain walls or just velocity). Therefore, if the domain walls' velocity is assumed to come from the Lorentz boost, the noncanonical kinetic component of domain walls $\dot{\phi}^4 \leq 0$. Since $\dot{\phi}^4\propto v^4a^4$, the scale factor $a$ dynamically changes the kinetic energy of the domain walls according to the observer, even though the domain walls' velocity is constant. For velocity $v \neq 0$, the absolute kinetic component of the domain walls increases with the expansion of the universe, following $a^4$. Therefore, the kinetic energy of the domain walls becomes more significant in the late epoch than in the early epoch. However, the last expression of Equation~(\ref{eq:energi-kinetik-kink-non-kanonik}) reveals that a domain wall may resemble a phantom field, i.e., the kinetic component is negative. To date, phantom stability is still an open topic in particle physics and cosmology. We will analyze the stability conditions for the noncanonical domain walls in the next subsection.

We have shown that the family of dynamic solutions obtained by performing a Lorentz boost can represent the phantom field properties of noncanonical domain walls for $v \neq 0$. Although we still owe a physical explanation for
the cause of the boost, the solution is mathematically acceptable considering the Lorentz invariant property of the Lagrangian. In addition, performing a Lorentz boost on the Bogomolnyi equation is a possible method to eliminate the imaginary parts in the energy density and pressure in Equation~(\ref{eq:rho-dan-P}).

\subsection{Unification of dark matter and dark energy}

Cosmological observations show that around $96\%$ of the universe's energy budget is in the form of dark matter and dark energy, which are still poorly understood. The nature of the two entities is different. On a large scales dark matter muffles the expansion of the universe, while on the other hand, dark energy boosts and accelerates cosmic dynamics. On the smaller scale, the flattening of the galaxy rotation curves shows that dark matter interacts with baryonic matter through gravitational fields \cite{rubin_1978}. Meanwhile, small-scale observations indicate that dark energy does not affect the dynamics of the bodies in the solar system, as inferred from the null results in observations of the fifth force effects. The Cassini mission reports that the post-Newtonian parameter in the solar system scale is $\lvert \gamma -1 \rvert < 2.3 \times 10^{-5}$ \cite{hoyle_2004}, meaning that the contribution of a fifth force is effectively negligible.

Notwithstanding that the empirical properties of the two are very different, several authors have attempted to unify dark matter and dark energy and perceive them as the same entity. Two examples of the proposed unification models of dark energy - dark matter are generalized Chaplygin gas and K-essence dark energy - dark matter (see \cite{amendola_tsujikawa_2010}). The philosophical reason for this unification is our lack of understanding of the true nature of the two entities. Our presumptions regarding the distinction between the two may be incorrect. In addition, a more practical reason is that we want to describe them through only one theory or model. In this work, we show that the noncanonical domain walls in certain circumstances can behave like dark energy and, in others, like cosmological dark matter.

Plugging Equation~(\ref{eq:energi-kinetik-kink-non-kanonik}) into Equation~(\ref{eq:rho-dan-P}), the energy density of the domain walls reads

\begin{equation}
    \rho_{dw}=4\left\langle V \right\rangle + 2 \left\langle v^2 \right\rangle \left\langle V \right\rangle a^2 - \left\langle v^4 \right\rangle \left\langle V \right\rangle a^4,
    \label{eq: rho(v,V}
\end{equation}

\noindent
and its pressure is

\begin{equation}
    P_{dw}=-6\left\langle V \right\rangle a^2 + \frac{10}{3}\left\langle v^2 \right\rangle \left\langle V \right\rangle a^4 - \frac{1}{3}\left\langle v^4 \right\rangle \left\langle V \right\rangle a^6.
    \label{eq:P(v,V)}
\end{equation}

\noindent
Therefore, if the network of the walls obeys the equation of state (EoS) $P_{dw}=w_{dw}\rho_{dw}$, then its EoS parameter reads

\begin{equation}
    w_{dw} = \frac{-6 a^2 + \frac{10}{3} v^2 a^4 - \frac{1}{3} v^4 a^6}{4 + 2 v^2 a^2 - v^4 a^4},
    \label{eq:w_DW(a,v}
\end{equation}

\noindent
where for convenience, the notation $\langle v \rangle$ has been replaced by $v$ without changing the meaning, namely, the velocity of the domain walls according to the observer's rest frame, in units of $c$. Equation~(\ref{eq:w_DW(a,v}) shows that the EoS parameter of the domain walls at a scale factor $a$ depends only on the magnitude of its velocity, as seen by the observer.

\subsubsection{Nonrelativistic and relativistic domain walls}

For nonrelativistic domain walls ($v\approx0$), we see from Equation~(\ref{eq:w_DW(a,v}) that the EoS parameter  $w_{dw}(a)\approx-\frac{3}{2}a^2$. In the distant past, when $a\ll1$,  $w_{dw}\approx0$. The density of the domain walls in
this era dominates over the pressure, i.e., $\rho_{dw}=\textrm{constant}>0$ and $P_{dw}\approx0$  (see Equation~(\ref{eq: rho(v,V}) and ~(\ref{eq:P(v,V)})), and the domain walls resemble cold dark matter. In the late epoch, when $a\rightarrow{1}$, the EoS parameter goes to  $w_{dw}\approx-1.5$. In this regime, the negative pressure of the walls dominates its density, resulting in a repulsive force. This emergent force opposes gravitational attraction and causes cosmic acceleration. In this epoch, the domain walls act as dark energy. Along the evolutionary path of the universe, domain walls with $v\approx0$ possess values of $w_{dw}$, which interpolate $0$ and $-1.5$ through Equation~(\ref{eq:w_DW(a,v}) with vanishing $v$.

For relativistic domain walls ($v \approx 1$), in the distant past ($a\ll1$), the EoS parameter  $w_{dw}$ approaches 0. Similar to the nonrelativistic case, the density of the domain walls in this era dominates the pressure, so the domain walls behave more like dark matter. Since $v\approx1$, the proper type of the resembled dark matter is hot or relativistic dark matter. In contrast, in the late epoch as $a$ approaches $1$, the negative pressure of the domain walls dominates the density. This condition evokes a repulsive force that generates cosmic acceleration with $w_{dw}\approx-0.6$, and the domain walls act as dark energy. Along the evolutionary path of the universe, domain walls with $v\approx{1}$ have EoS parameter values that interpolate $w_{dw}\approx0$ and $w_{dw}\approx-0.6$ via Equation~(\ref{eq:w_DW(a,v}) for $v$ approaching $1$.

\subsubsection{Phantom state and phantom zone}

\begin{figure}
\centering
    \includegraphics[width=1.00\columnwidth]{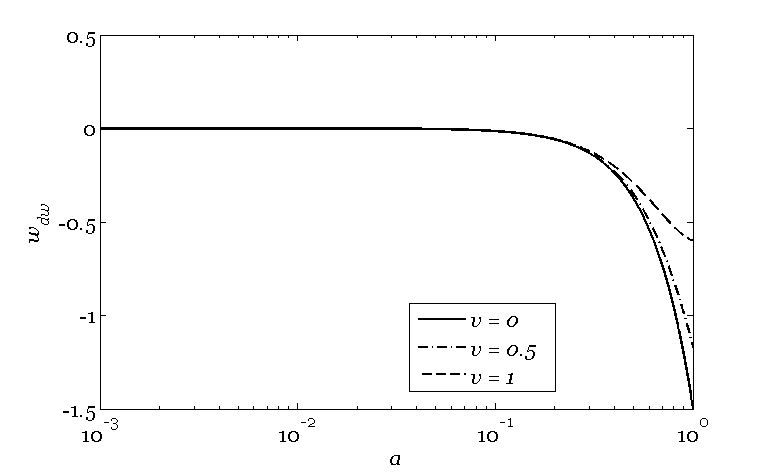}
    \caption{Evolution of the EoS parameter of noncanonical domain walls for several values of velocity $v$}
    \label{fig:figure1}
\end{figure}

Figure~\ref{fig:figure1} shows the evolution of the EoS parameter along with the expansion history of the universe for several velocity values of the noncanonical domain walls. Domain walls with $0<v<1$ have $w_{dw}(a,v)$, which interpolates $w_{dw}(a,v=0)$ and $w_{dw}(a,v=1)$. However, note that domain walls with any $v$ will converge to $w_{dw}\approx 0$ when $a\rightarrow{0}$. In other words, all noncanonical domain walls are initially created as dark matter. As the universe expands, the pressure of the domain walls becomes increasingly negative and slowly takes over the corresponding density, and $w_{dw}$ becomes increasingly negative as well. The higher the velocity of the domain walls, the slower the decrease of $w_{dw}$ with the expansion of the universe. Clearly, some domain walls fall into the phantom zone $(w_{dw}<-1)$ at the late epoch. Domain walls that enter the phantom zone are those whose velocities are smaller than a characteristic velocity $v_{ph}$ that satisfies

\begin{equation}
    \frac{-6 a^2 + \frac{10}{3} v_{ph}^2 a^4 - \frac{1}{3} v_{ph}^4 a^6}{4 + 2 v_{ph}^2 a^2 - v_{ph}^4 a^4} = w_{dw}\left(a,v_{ph}\right)=-1.
\end{equation}

\noindent
Today, $a=1$, so that $v_{ph}=0.65$. Therefore, domain walls with $w_{dw}=-1$ today have $v=v_{ph}$, and domain walls with $v<v_{ph}$ today are entering the phantom zone.

Equation (\ref{eq:energi-kinetik-kink-non-kanonik}) clearly shows that domain walls with Lagrangian~(\ref{eq:L2}) have negative kinetic energy unless they have zero velocity. However, Figure~\ref{fig:figure1} confirms that domain walls with $v=0$ are in the phantom zone ($w_{dw}(a=1)=-1.5)$ today. Since the domain walls have constant velocities, those that are frozen ($v=0$) today were also frozen in the past. Therefore, they never possess negative kinetic energy along their trajectory. A state in which the kinetic energy of the domain walls is negative (ghost) is called a phantom state. Thus, a phantom state and a phantom zone (a region in which the domain walls have $w_{dw}<-1$) refer to two different ideas.

At a specific moment in the late epoch, domain walls with $v=0$ will enter the phantom zone while retaining positive kinetic energy. Although they are not phantoms, they will eventually reach the phantom zone. In addition, domain walls with $0<v<v_{ph}$ are always in the phantom state and eventually enter the phantom zone. In contrast, domain walls whose velocity $v>v_{ph}$ along their evolutionary path will never enter the phantom zone, even if they are perpetually in a phantom state.

\subsection{Stability conditions}

One possible solution to avoid negative kinetic energy in Equation (\ref{eq:energi-kinetik-kink-non-kanonik}) is by assuming a vanishing velocity for the domain walls, i.e., it does not move ($v=0$) according to an observer in the $O(t,x,y,z)$ frame. We now investigate the stability of the noncanonical domain walls under a small perturbation $\delta\phi(x,t)$ around the field $\phi_0(x,t)$. At the linear perturbation level, the field $\phi(x,t)$ becomes

\begin{equation}
    \phi(t,x)=\phi_0(t,x) + \delta\phi(t,x).
\end{equation}

\noindent
We rewrite Lagrangian~(\ref{eq:L2}) in the form of

\begin{equation}
    \mathcal{L}(K,\phi) = K^2 - V(\phi); \quad K \equiv \frac{1}{2}g^{\mu\nu}\partial_{\mu}\phi\partial_{\nu}\phi;
\end{equation}

\noindent
and express the Hamiltonian in the form of

\begin{equation}
    \mathscr{H} = K^2+V(\phi).
    \label{eq:hamiltonian}
\end{equation}

Up to its second term, the perturbation of Hamiltonian~(\ref{eq:hamiltonian}) is given by \cite{amendola_tsujikawa_2010}

\begin{equation}
\begin{split}
    \delta\mathscr{H} = \left( \partial_K\mathcal{L} + 2K\partial_K\left(\partial_K\mathcal{L}\right) \right) \frac{\left(\partial_t\delta\phi\right)^2}{2}  + \frac{1}{2}\partial_K\mathcal{L} g^{\mu\nu}\partial_{\mu}\delta\phi\partial_{\nu}\delta\phi
    - \partial_{\phi}\left(\partial_{\phi}\mathcal{L}\right)\frac{\left(\delta\phi\right)^2}{2}.
    \label{eq:gangguan hamiltonian}
\end{split}
\end{equation}

\noindent
The expansion of the perturbed Hamiltonian above will be either positive or zero if the following conditions are satisfied:

\begin{equation}
    \begin{split}
        \xi_1 \equiv \partial_K\mathcal{L} + 2K\partial_K\left(\partial_K\mathcal{L}\right) \geq 0;\\
        \xi_2 \equiv \partial_K\mathcal{L} \geq 0;\\
        \xi_3 \equiv - \partial_{\phi}\left(\partial_{\phi}\mathcal{L}\right) \geq 0.
        \label{eq:syarat stabil kuantum}
    \end{split}
\end{equation}

\noindent
Let us explore Equation~(\ref{eq:gangguan hamiltonian}) in more detail. The stability conditions of the domain walls depend both on the shape of the kinetic and the potential terms. For the kinetic term, the stability conditions are $\xi_1 \geq 0$ and $\xi_2 \geq 0$, while $\xi_3 \geq 0$ is the stability condition for the potential term.

The Lagrangian in Equation~(\ref{eq:L2}) gives $\xi_1=6K \geq 0$ and $\xi_2=2K \geq 0$. Comparing these with Equation~(\ref{eq:energi-kinetik-kink-non-kanonik}), we find that these two conditions are true for $v=0$. This is expected from the discussion in subsection \ref{subsec:phantom-properties} regarding the kinetic energy of the domain walls. Meanwhile, the third stability condition, namely, the stability of the potential component, can be written as

\begin{equation}
    \xi_3 = -\partial_{\phi} \left( -\frac{\partial V(\phi)}{\partial\phi} \right) = \frac{\partial^2V(\phi)}{\partial\phi^2}
\end{equation}

\noindent
for any potential $V(\phi)$. This third stability condition is met by domain walls whose second derivative of the potential with respect to the scalar field is either positive or zero.

In addition to the quantum mechanics stability, we also need to investigate the stability conditions of the domain walls according to classical mechanics and special relativity. Regarding classical mechanics, the perturbed Lagrangian~(\ref{eq:L2}) is stable if the sound speed (see \cite{amendola_tsujikawa_2010})

\begin{equation}
    c_s^2 \equiv \frac{\partial_K\mathcal{L}}{\partial_K\mathcal{L}+2K\partial_K\left(\partial_K\mathcal{L}\right)} > 0
\end{equation}

\noindent
is fulfilled. Domain walls with $v=0$ satisfy this condition because they have $c_s^2=\frac{1}{3}>0$. They also obey the special relativity conditions because $c_s^2<1$ and hence do not violate the causality principle.

We have obtained the stability conditions and conclude that the only stable dynamic solution for the noncanonical domain walls is $v=0$. Therefore, a noncanonical domain wall occupying its quantum ground state is a domain wall with Lagrangian~(\ref{eq:L2}) that physically does not suffer from quantum instability. In the canonical formalism, a domain wall with zero velocity is called a frustrated domain wall \cite{friedland_2002,fabris_2003,lazanu_2015,peyravi_2016}. However, in the noncanonical formalism, for distinction and brevity, we call it a freezing domain wall.

If the topological defects of the early universe generate domain walls with Lagrangian (\ref{eq:L2}) and velocities $0\leq v \leq 1$, the domain walls must immediately be frozen after being created to evade instability; otherwise, they are unstable and physically never exist. However, we still have no idea whether the unstable domain walls decay into other entities or never really existed.

\subsection{Cosmological background dynamics}

Let us suppose the universe contains radiation, matter, and domain walls as dark energy of densities $\rho_r$, $\rho_m$, and $\rho_{dw}$, respectively. The background dynamics of the universe are described well by the Friedmann equation, acceleration equation, continuity equation for perfect fluid, and EoS of each constituent which obeys $P=w\rho$. Assuming a flat universe, we obtain

\begin{equation}
    \left(\frac{\dot{a}}{a}\right)^2 = \frac{8\pi G}{3} \rho_T;
    \label{eq:persamaan Friedmann}
\end{equation}

\vspace{0mm}

\begin{equation}
    \frac{\ddot{a}}{a} = - \frac{4\pi G}{3}\left(\rho_T + 3P_T\right);
    \label{eq:persamaan percepatan}
\end{equation}

\vspace{0mm}

\begin{equation}
    \dot{\rho}_X + 3H\left(\rho_X+P_X\right)=0.
    \label{eq:persamaan kontinuitas}
\end{equation}

\noindent
Respectively, the subscripts $T$ and $X$ denote the total contribution of all components and each particular component. The density parameter of component $X$ is defined as $\Omega_X(a) \equiv \rho_X(a)/\rho_{crit}(a)$, where $\rho_{crit}(a) \equiv 8\pi G/3H^2$. Thus, for the postinflation era to date, the flat universe satisfies the condition
\begin{equation}
    \Omega_T = \Omega_r(a) + \Omega_m(a) + \Omega_{dw}(a) = 1.
\end{equation}

Globally, each component of the universe evolves independently. The omission of couplings between different components is acceptable in the study of the expansion history at the background level.
Therefore, we can solve continuity Equation~(\ref{eq:persamaan kontinuitas}) for each constituent independently, i.e.,

\begin{equation}
    \rho_r = \rho_{r,0}a^{-4}=\rho_{r,0}\left(1+z\right)^4
    \label{eq:rho radiasi}
\end{equation}
\noindent
for radiation,
\begin{equation}
    \rho_m = \rho_{m,0}a^{-3}=\rho_{m,0}\left(1+z\right)^3
    \label{eq:rho materi}
\end{equation}
\noindent
for matter, and
\begin{equation}
    \rho_{dw} = \rho_{dw,0} \times \textrm{exp}
        \left[ \int_{0}^{z} \frac{3\left(1+w_{dw}(z')\right)}{1+z'} \,dz' \right].
        \label{eq:rho domain wall}
\end{equation}

\noindent
for the domain walls, where the subscript $0$ refers to today's value.

As the universe expands, the energy density of radiation decreases faster than that of matter. Equation~(\ref{eq:rho domain wall}) shows that all the time, the density of the domain walls depends globally on the EoS parameter $w_{dw}$, which is dependent only on the velocity measured by the observer. Considering the result of the stability tests in the previous section, we investigate the dynamics of the background universe with the freezing domain walls ($v=0$) as dark energy.
Equation~(\ref{eq:w_DW(a,v}) shows that the freezing domain walls satisfy $w_{dw}(a)=-1.5a^2$, and today they are in the phantom zone. Recall, however, that their energy states are stable.

The solution of the integral factor in Equation~(\ref{eq:rho domain wall}) is

\small
\begin{equation*}
    \int_{0}^{z} \frac{3\left(1+w_{dw}(z')\right)}{1+z'} \,dz' = 3 \left[ 
    \textrm{ln}\left(1+z\right)+\frac{3}{4z^2+8z+4}-\frac{3}{4} \right],
\end{equation*}
\normalsize

\noindent
so that Equation~(\ref{eq:rho domain wall}) becomes

\begin{equation}
    \rho_{dw} = \rho_{dw,0}(1+z)^{3} \textrm{exp} \left( -\frac{9}{4}\frac{z(z+2)}{(1+z)^{2}} \right).
    \label{eq:rho domain wall final}
\end{equation}

\noindent
For large $z$, the exponential factor of Equation~(\ref{eq:rho domain wall final}) is small and practically constant; hence, the evolution of the energy density of the domain walls resembles that of  matter, namely, $\rho_{dw}\propto(1+z)^3$. Since observations tightly constrain the abundance of baryons, the domain walls should not be baryonic matter but act more like dark matter. In late times (small $z$), the cubic factor approaches 1, and the domain walls slowly deviate from the matter properties and turn into dark energy.

\begin{figure}
\centering
    \includegraphics[width=1.00\columnwidth]{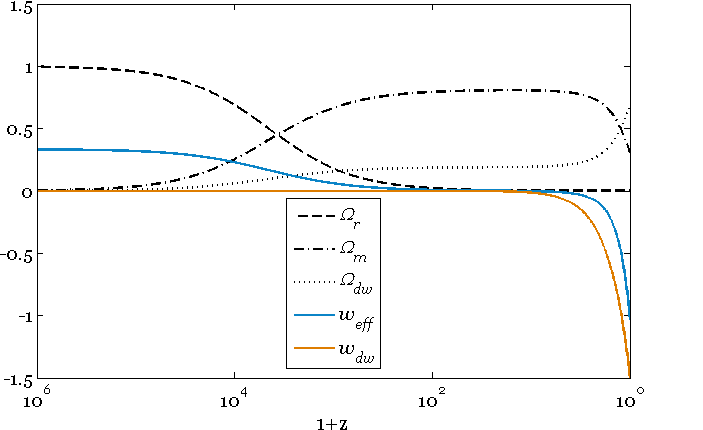}
    \caption{Cosmic evolutionary paths of $\Omega_r$, $\Omega_m$, $\Omega_{dw}$, $w_{eff}$ and $w_{dw}$ for a universe with freezing domain walls $(v=0)$.}
    \label{fig:figure3}
\end{figure}

Figure~\ref{fig:figure3} shows the cosmic density evolution of the three different components of the universe (radiation, matter, and freezing domain walls) for the input parameters $\Omega_{m,0}=0.31$ \cite{planck_2020} and $\Omega_{r,0}=8.4 \times10^{-5}$ (see \cite{ryden_2017}). Figure~\ref{fig:figure3} also shows that the noncanonical freezing domain walls generate late-time cosmic acceleration. In early times, when radiation drives the dynamics of the universe, $\Omega_{dw}$ is extremely small ($\sim 10^{-4}-10^{-3}$) with $w_{dw} \approx 0$. During that moment, the domain walls behave similarly to matter. As the universe enters the transition era of matter-radiation, $\Omega_r$ decreases with increasing $\Omega_m$. As $\Omega_m$ increases, $\Omega_{dw}$ also increases, while $w_{dw}$ remains unchanged. The matter characteristics of the domain walls are increasingly noticeable as the universe enters the matter-dominated era. Over a long period, matter and domain walls share portions of energy, so during this era $\Omega_m\leq 1$. However, since domain walls behave like matter, they contribute to matter density. Hence, the effective matter density parameter is $\Omega_m + \Omega_{dw}$. The two constituents take part in gravitational interactions. Furthermore, domain walls may contribute to the creation and amplification (or deamplification) of overdense regions. During that period, the freezing domain walls act as cold or non-relativistic dark matter. Therefore, we may expect a different scenario of structure formation compared to that of the $\Lambda$CDM scenario.

At late times, domain walls start to deviate from their matter properties, the EoS parameter gradually becomes negative, and their dark energy properties slowly emerge. According to the freezing domain wall scenario, the transition from the matter-dominated era to the domain wall era occurs at $z \approx 0.24$ or equivalently at $a \approx 0.80$.  Hence, the model predicts that the time of matter-dark energy equality is slightly later than that of $\Lambda$CDM ($a_{m\Lambda} \approx 0.75$).  In the later times, when the domain walls dominate, massive expansion occurs since $w_{dw}$ becomes more negative and subsequently enters the phantom zone. Today, domain walls are playing the role of dark energy with $w_{dw,0} \approx -1.5$ and an effective EoS parameter that includes all components $w_{eff,0} \approx -1.03$. We show in Table \ref{tbl:w0} that our model predicts a value of $w_{dw,0}$, which is in good agreement with the CMB Planck measurements for temperature fluctuation, polarization, lensing data, and their combinations \cite{planck_2020}. However, the value no longer fits when BAO data are jointly analyzed with the CMB data.

\begin{table}[tbp]
\centering
\begin{tabular}{|c|c|}
\hline
\textbf{Cosmological model/dataset} & \textbf{$w_{de,0}$}\\
\hline
Our model & \small $-1.5$\\
\textit{TT+lowE} & \small $-1.56^{+0.60}_{-0.48}$\\
\textit{TT,TE,EE+lowE} & \small $-1.58^{+0.52}_{-0.41}$\\
\textit{TT,TE,EE+lowE+lensing} & \small $-1.57^{+0.50}_{-0.40}$\\
\textit{TT,TE,EE+lowE+lensing+BAO} & \small $-1.04^{+0.1}_{-0.1}$\\
\hline
\end{tabular}
\caption{The current value of the dark energy EoS parameter following our model and the results of Planck for various datasets \cite{planck_2020}}
\label{tbl:w0} % spaces are big no-no withing labels
\end{table}

\section{Inhomogeneous universe with noncanonical domain walls}

\subsection{Analytical study}
Here, we employ linear cosmological perturbation theory. Any quantity or variable describing the universe can be expanded up to linear terms as follows:

\begin{equation}
    f_{\mu\nu}=\bar{f}_{\mu\nu}+\delta f_{\mu\nu}.
\end{equation}

\noindent
Since gravitational force is the dominant force acting on cosmological scales, we investigate the perturbation dynamics by expanding Einstein field equations into linear terms. The perturbation components for the Einstein field equations are

\begin{equation}
    \delta R_{\mu\nu}-\frac{1}{2} \left( \delta g_{\mu\nu}R+g_{\mu\nu}\delta R \right) = 8\pi G \delta T_{\mu\nu}.
    \label{eq:gangguan EFE}
\end{equation}

\noindent
Assuming the energy of components in the universe obeys the energy-momentum tensor

\begin{equation}
    T_{\mu\nu}=\left(\rho+P\right)u_{\mu}u_{\nu}+P g_{\mu\nu},
\end{equation}

\noindent
all components in the universe experience energy-momentum perturbations

\small
\begin{equation}
\begin{split}
    \delta T_{\nu}^{\mu} =
    \rho \left[ u^{\mu}u_{\nu}\delta\left(1+c_s^2\right)+\left(u^{\mu}\delta u_{\nu}+\delta u^{\mu}u_{\nu}\right)\left(1+w\right)+c_s^2\delta\delta_{\mu}^{\nu} \right],
\end{split}
\end{equation}
\normalsize

\noindent
with

\begin{equation}
    c_s^2\equiv\frac{\delta P}{\delta\rho}=\frac{dP}{d\rho}=\frac{\dot{P}}{\dot{\rho}}, \quad \textrm{and} \quad \delta\equiv\frac{\delta\rho}{\bar{\rho}}=\frac{\rho-\bar{\rho}}{\bar{\rho}},
\end{equation}

\noindent
respectively express the sound speed and the density contrast, and $u^{\mu}$ is the four-velocity with normalization $u_{\mu}u^{\mu}=-1$. To obtain a set of perturbation equations, we first need to select the gauge. For convenience, we apply the conformal Newtonian or longitudinal gauge in our analytical study, while the synchronous gauge is used in the numerical study for the reasons of computational stability.

Using the conformal Newtonian or longitudinal gauge, we examine the properties of domain walls when acting as dark matter and as dark energy through their influence on the behavior of matter perturbations. In this subsection, we follow the procedure and derivation in \cite{amendola_tsujikawa_2010} for the general dark energy models. We first ignore the presence of radiation and other relativistic matter to simplify the analysis. We shall involve the contribution of radiation and neutrinos in the next subsection.

In the conformal time $\eta \equiv \int a^{-1}dt$, the line element of perturbed spacetime following the longitudinal gauge is

\begin{equation}
    ds^2=a^2(\eta)\left[ -\left(1+2\Psi\right)d\eta^2 + \left(1+2\Phi\right)\delta_{ij}dx^i dx^j \right],
    \label{eq:elemen garis longitudinal gauge}
\end{equation}

\noindent
where $\Psi(r,a)=\Psi(a)e^{i\vec{k} \cdot r}$ and $\Phi(r,a)=\Phi(a)e^{i\vec{k} \cdot r}$ are Bar-deen potentials, and $\delta_{ij}$ is the Kronecker delta. Expanding the field Equation (\ref{eq:gangguan EFE}) for the line element (\ref{eq:elemen garis longitudinal gauge}), we obtain

\begin{equation}
    3\mathcal{H} \left( \mathcal{H}\Psi-\frac{\partial\Phi}{\partial\eta} \right) + \nabla^2\Phi = -4\pi G a^2 \delta\rho;
    \label{eq:gangguan EFE-1}
\end{equation}

\vspace{0mm}

\begin{equation}
    \nabla^2 \left( \frac{\partial\Phi}{\partial\eta}-\mathcal{H}\Psi \right)=4\pi G a^2\left(1+w\right)\rho \hat{\theta};
    \label{eq:gangguan EFE-2}
\end{equation}

\vspace{0mm}

\begin{equation}
    \Psi=-\Phi;
    \label{eq:gangguan EFE-3}
\end{equation}

\vspace{0mm}

\begin{equation}
    \frac{\partial^2\Phi}{\partial\eta^2}+2\mathcal{H}\frac{\partial\Phi}{\partial\eta}-\mathcal{H}\frac{\partial\Psi}{\partial\eta}-\left(\mathcal{H}^2+2\frac{\partial\mathcal{H}}{\partial\eta}\right)\Psi=-4\pi G a^2 c_s^2\delta\rho,
    \label{eq:gangguan EFE-4}
\end{equation}

\noindent
where $\mathcal{H}\equiv aH$ is the conformal Hubble parameter, $\hat{\theta} \equiv \nabla_i v^i$ is the velocity divergence, and $v^i = a\frac{dx^i}{dt}$ is the peculiar velocity in the comoving frame. The four equations (\ref{eq:gangguan EFE-1}, \ref{eq:gangguan EFE-2}, \ref{eq:gangguan EFE-3}, and \ref{eq:gangguan EFE-4}) describe the relations between density perturbation and velocity divergence with Bardeen potentials. In other words, inhomogeneity of the universe might occur either through the presence of density perturbations or velocity divergences. However, these four equations are insufficient to describe the overall formation and evolution of inhomogeneities.

Up to this point, we have not included the perturbations of the mass-energy components. Another requisite is the fulfilment of energy-momentum conservation at the background as well as at perturbation levels. At the background level, the energy-momentum conservation, i.e., $T_{\nu;\mu}^\mu=0$ gives the continuity equation $\frac{\partial\rho}{\partial\eta}+3\mathcal{ H}\left(\rho+P\right)=0$. Parsing $\delta T_{\nu;\mu}^\mu=0$ for $\nu=0$, applying the continuity equation, and expressing it in time coordinate $\tau=\ln a$, we find

\begin{equation}
    \frac{\partial\delta_X}{\partial\tau}=3\left(w_X-c_{s,X}^2\right)\delta_X-\left(\theta_X+3\frac{\partial\Phi}{\partial\tau}\right)\left(1+w_X\right),
    \label{eq:pers. gangguan kerapatan delta longitudinal}
\end{equation}

\noindent
where the subscript $X$ refers to matter and domain walls, and for convenience, we have defined the normalized velocity divergence

\begin{equation}
    \theta \equiv \frac{\hat{\theta}}{\mathcal{H}}.
    \label{eq:definisi theta}
\end{equation}

\noindent
Meanwhile, from $\delta T_{\nu;\mu}^\mu=0$ for $\nu=i$, we find

\begin{equation}
\begin{split}
    \frac{\partial\theta_X}{\partial\tau} =
    \left( 3w_X-1-\frac{1}{1+w_X}\frac{\partial w_X}{\partial\tau}-\frac{1}{\mathcal{H}} \frac{\partial\mathcal{H}}{\partial\tau} \right)\theta_X
    + \frac{c_{s,X}^2}{\hat{\lambda}^2\left(1+w_X\right)}\delta_X - \frac{\Phi}{\hat{\lambda}^2},
    \label{eq:pers. divergensi kecepatan theta longitudinal}
\end{split}
\end{equation}

\noindent
where we have introduced the quantity

\begin{equation}
    \hat{\lambda} \equiv \frac{\mathcal{H}}{k} = \frac{aH}{k},
    \label{eq:lambda-topi}
\end{equation}

\noindent
and $k$ is the perturbation wavenumber. Equations (\ref{eq:pers. gangguan kerapatan delta longitudinal}) and (\ref{eq:pers. divergensi kecepatan theta longitudinal}) are Boltzmann equations for matter in the presence of domain walls as dark energy and $w_X \neq -1$.  Following these two equations, the homogeneity of mass distribution in an area may break down due to several reasons. First, some regions have densities that continuously change relative to others $\left( \frac{\partial \delta_X}{\partial \tau} > {0} \right)$; second, some regions have different velocities that change continuously relative to others $\left( \frac{\partial \theta_X}{\partial \tau} > {0}\right)$; third, a combination of the two cases. If none of these reasons are met, the homogeneity (or inhomogeneity) of the universe remains the same as the initial conditions or decreases.

For $w_X \approx 0$, Equation (\ref{eq:pers. gangguan kerapatan delta longitudinal}) implies that the evolution of a density contrast depends on the magnitude of the velocity divergence of the mass-energy distribution $(\theta_X)$ and the change in the gravitational potential $\left( \frac{\partial \Phi}{\partial \tau} \right)$. However, from Equation (\ref{eq:pers. divergensi kecepatan theta longitudinal}),  we know that the velocity divergence $(\theta_X)$ itself depends on the gravitational potential $\Phi$. Therefore, the growth of inhomogeneity would be different for constant and varying gravitational potentials $\Phi$.

In the case of a constant gravitational potential, gravity does not directly change the density contrast. However, it does change the velocity divergence. The nonzero velocity divergence increases the density contrast,  gradually enhancing inhomogeneity in the universe. In contrast, if the gravitational potential varies with time, density contrast and velocity divergence changes may co-occur. Since velocity divergence itself induces  density contrast,  inhomogeneity would change more rapidly in a nonconstant gravitational potential than in a constant gravitational potential.

Thus far, we have reviewed the density contrast and velocity divergence of each component, starting from the perturbed energy-momentum tensor and conservation law. We shall now examine interconnections between these perturbation variables, starting from Einstein field equations, to find the total density contrast and total velocity divergence. We can express Equations (\ref{eq:gangguan EFE-1}) and (\ref{eq:gangguan EFE-2}) in the form of

\begin{equation}
    \Phi = 3\hat{\lambda}^2 \left( \frac{1}{2}\delta_T-\Phi-\frac{d\Phi}{d\tau} \right)
    \label{eq:potensial dari EFE-1}
\end{equation}

\noindent
and

\begin{equation}
    \frac{d\Phi}{d\tau} = -\Phi - \frac{3}{2}\hat{\lambda}^2\left(1+w_{eff}\right)\theta_T,
    \label{eq:potensial dari EFE-2}
\end{equation}

\noindent
where we have used Equation (\ref{eq:gangguan EFE-3}), $3\mathcal{H}=8\pi G a^2 \rho_T$, and the definitions $\delta_T \equiv \sum_i \Omega_i \delta_i$, $\theta_T \equiv \sum_i \frac{1+w_i}{1+w_{eff}}\Omega_i \theta_i$, and $w_{eff} \equiv \sum_i \Omega_i w_i$. Combining Equations (\ref{eq:potensial dari EFE-1}) and (\ref{eq:potensial dari EFE-2}), we easily obtain

\begin{equation}
    \Phi = \frac{3}{2} \hat{\lambda}^2 \left[ \delta_T+3\hat{\lambda}^2\left(1+w_{eff}\right)\theta_T \right],
    \label{eq:potensial dari EFE-1 dan EFE-2}
\end{equation}

\noindent
that describes gravitational potential in terms of the total density contrast and total velocity divergence. For a perturbed universe containing matter and domain walls, the sound speed of perturbation reads

\begin{align}
    c_{s,T}^2 &= \frac{\sum_i c_{s,i}^2\Omega_i \delta_i}{\sum_i \Omega_i \delta_i} = c_{s,dw}^2 \left( 1-\frac{\Omega_m \delta_m}{\delta_T} \right) \\
    &= c_{s,dw}^2 \left[ 1-\frac{\Omega_m \delta_m}{2\frac{d\Phi}{d\tau}+2\Phi \left( 1+\frac{1}{3\hat{\lambda}^2} \right) } \right],
    \label{eq:CsDW}
\end{align}

\noindent
where we have used Equation (\ref{eq:potensial dari EFE-1}).

Combining the two first-order differential Equations (\ref{eq:pers. gangguan kerapatan delta longitudinal}) and (\ref{eq:pers. divergensi kecepatan theta longitudinal}), we have a second-order perturbation equation. Differentiating Equation (\ref{eq:pers. gangguan kerapatan delta longitudinal}) and using Equation (\ref{eq:pers. divergensi kecepatan theta longitudinal}), the second-order perturbation equation is \cite{amendola_tsujikawa_2010}

\begin{equation}
    \frac{d^2\delta_X}{d\tau^2}+a_1\frac{d\delta_X}{d\tau}+a_0\delta_X=b_0\Phi+b_1\frac{d\Phi}{d\tau} + b_2\frac{d^2\Phi}{d\tau^2},
    \label{eq:pers. gangguan orde-dua}
\end{equation}

\noindent
where

\begin{equation}
\begin{aligned}
    a_0 &\equiv \frac{1}{2} \left[ c_{s,X}^2\left(3+2\hat{\lambda}^{-2}-18w_X-9w_{eff}\right) \right. \\
    & \left. + 3w_X\left(-1+6w_X+3w_{eff}\right) + 6\left( \frac{d\left[c_{s,X}^2\right]}{d\tau}-\frac{dw_X}{d\tau} \right) \right];\\
    a_1 &\equiv \frac{1}{2}\left(1+6c_{s,X}^2-12w_X-3w_{eff}\right);\\
    b_0 &\equiv \hat{\lambda}^{-2}\left(1+w_X\right);\\
    b_1 &\equiv -\frac{3}{2}\left[ 2\frac{dw_X}{d\tau}+\left(1+w_X\right)\left(1-6w_X-3w_{eff}\right) \right];\\
    b_2 &\equiv -3\left(1+w_X\right).
    \label{eq:koefisien pers. gangguan orde-dua}
\end{aligned}
\end{equation}

\noindent

\subsubsection{Matter perturbation}

We now consider a case in which gravitational matter instability occurs in the presence of the domain walls. In the case of pressureless matter perturbation ($w_X=w_m\approx 0$), Equation (\ref{eq:pers. gangguan orde-dua}) reads

\small
\begin{equation}
    \frac{d^2\delta_m}{d\tau^2}+\frac{\left(1-3w_{eff}\right)}{2}\frac{d\delta_m}{d\tau}=\frac{\Phi}{\hat{\lambda}^{2}}-\frac{3\left(1-3w_{eff}\right)}{2}\frac{d\Phi}{d\tau}-3\frac{d^2\Phi}{d\tau^2}.
    \label{eq:pers. gangguan orde-dua materi}
\end{equation}
\normalsize

\noindent
On subhorizon scales  ($\hat{\lambda} \ll 1$), Equation (\ref{eq:pers. gangguan orde-dua materi}), (\ref{eq:potensial dari EFE-1 dan EFE-2}), and (\ref{eq:CsDW}) become

\begin{equation}
    \frac{d^2\delta_m}{d\tau^2}+\frac{1}{2}\left(1-3w_{eff}\right)\frac{d\delta_m}{d\tau}=\frac{\Phi}{\hat{\lambda}^2};
    \label{eq:pers. gangguan orde-dua materi subhorizon}
\end{equation}

\begin{equation}
    \Phi = \frac{3}{2}\hat{\lambda}^2\delta_T = \frac{3}{2}\hat{\lambda}^2\left(\Omega_m\delta_m + \Omega_{dw}\delta_{dw}\right);
    \label{eq:potensial dari EFE-1 dan EFE-2 subhorizon}
\end{equation}

\begin{equation}
    c_{s,T}^2 = c_{s,dw}^2 \left[ 1-\frac{3\hat{\lambda}^2\Omega_m \delta_m}{2\Phi} \right].
    \label{eq:CsDW subhorizon}
\end{equation}

\noindent
The three equations fully describe the evolution of matter density contrasts in the presence of domain walls.

If the domain walls are unclustered ($\delta_{dw}=0$), Equation (\ref{eq:potensial dari EFE-1 dan EFE-2 subhorizon}) becomes $\Phi = \frac{3}{2}\hat{\lambda}^2\Omega_m\delta_m$; hence, the sound speed in Equation (\ref{eq:CsDW subhorizon}) vanishes. During the matter-dominated era, domain walls act as dark matter, so $\Omega_m+\Omega_{dw}=1$, or $\Omega_m <  1$. Therefore, the total gravitational potential $\Phi$ is less than its value in the $\Lambda$CDM scenario. As gravitational potential affects density contrast through Equation (\ref{eq:pers. gangguan orde-dua materi subhorizon}), the resulting density contrast following this model is smaller than that of the $\Lambda$CDM model. Indeed, our numerical simulations confirm the results of this analytical study, as shown in the following subsection.

\subsubsection{Growth rate of large scale structures}

One of the parameters that characterizes different dark energy models is  the growth rate of structures $g$, defined as $g\equiv d\!\ln\delta_m/d\! \ln a=\Omega_m^{\gamma}$, where $\gamma$ denotes the index of the
growth parameter. Using this definition, the combination of Equations (\ref{eq:pers. gangguan orde-dua materi subhorizon}) and (\ref{eq:potensial dari EFE-1 dan EFE-2 subhorizon}) for $\delta_{dw}=0$ gives

\begin{equation}
    \frac{dg}{d\tau}+g^2+ \left[ \frac{1}{2}-\frac{3}{2}w_{dw}\left(1-\Omega_m\right) \right]g=\frac{3}{2}\Omega_m.
    \label{eq:pers. gangguan dalam faktor pertumbuhan}
\end{equation}

\noindent
Making use of the Friedmann equation $3H^2=8\pi G (\rho_m+\rho_{dw})$ jointly with continuity equation $\dot{\rho}_{dw}+3H(1+w_{dw})\rho_{dw}=0$, we obtain

\begin{equation}
    \frac{d\Omega_m}{d\tau}=3w_{dw}\left(1-\Omega_m\right)\Omega_m.
    \label{eq:hubungan Omega_m aksen dan w_DW}
\end{equation}

\noindent
Using the definition of $g$ and combining Equations (\ref{eq:pers. gangguan dalam faktor pertumbuhan}) and (\ref{eq:hubungan Omega_m aksen dan w_DW}), Wang et al. \cite{Wang_1998} obtained the following perturbation equation:

\begin{equation}
\begin{split}
    3w_{dw}\Omega_m\left(1-\Omega_m\right) \left(\ln \Omega_m\right) \frac{d\gamma}{d\Omega_m} - 3w_{dw} \left( \gamma - \frac{1}{2} \right)\Omega_m + 
    \Omega_m^{\gamma} - \frac{3}{2}\Omega_m^{1-\gamma} + 3w_{dw}\gamma - \frac{3}{2}w_{dw} + \frac{1}{2} = 0.
    \label{eq:per. gangguan dalam gamma}
\end{split}
\end{equation}

\noindent
During the matter-dominated era, $w_{dw}=0$, and the change in $\frac{dw_{dw}}{d\Omega_m}$ is extremely small, so $ \lvert \frac{dw_{dw}}{d\Omega_m} \rvert \ll \frac{1}{1-\Omega_m} $. An approximation for the solution of Equation (\ref{eq:per. gangguan dalam gamma}) is given in \cite{Wang_1998} and \cite{Gong_2009}, i.e.,

\begin{equation}
\begin{split}    
    \gamma = \frac{3\left(1-w_{dw}\right)}{5-6w_{dw}}
    +\frac{3\left(1-w_{dw}\right)\left(1-\frac{3}{2}w_{dw}\right)}{125\left(1-\frac{6}{5}w_{dw}\right)^3} \left(1-\Omega_m\right)
    \label{eq:gamma-dari-wang-dan-gong}.
\end{split}
\end{equation}

\begin{figure}
\centering
    \includegraphics[width=1.00\columnwidth]{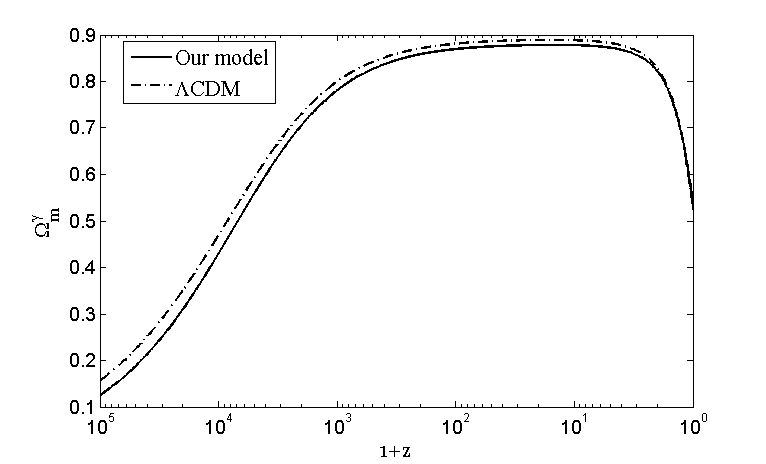}
   \caption{The growth rate of structures for the noncanonical domain walls (solid line) and $\Lambda$CDM (dash-dotted line) dark energy models following Equation (\ref{eq:gamma-dari-wang-dan-gong}) for $\Omega_{m,0}=0.31$}
    \label{fig:g_vs_z}
\end{figure}

Figure \ref{fig:g_vs_z} compares the structure growth behaviors for the noncanonical domain walls and $\Lambda$CDM dark
energy models across redshift. This result indicates that structures in the noncanonical domain wall scenario grow slightly slower than in the $\Lambda$CDM model, confirming our previous analysis.  As mentioned above,  domain walls act like dark matter during the matter-dominated regime and $\Omega_m < 1$. However, since domain walls do not cluster, they do not contribute to the gravitational instability at that time, so the density contrast is smaller than that of $\Lambda$CDM. At low redshifts, however, the two models approximately coincide. This means that the effect of the noncanonical domain walls on the formation and evolution of large-scale structures is not much different from that of the cosmological constant. We compare the growth parameters $g$ and the index of growth parameters $\gamma$ for the noncanonical domain wall and  $\Lambda$CDM models at redshift $z\approx 1$ in Table \ref{tbl:g_dan_gamma}.

\begin{table}
\centering
\begin{tabular}{ |p{4cm}|p{1cm}|p{1cm}| }
\hline
%\multicolumn{3}{|c|}{$z\approx 1$}\\ %z=10198
%\midrule
\textbf{Cosmological model} & $\gamma$ & $g$\\
\hline
%This model & $0,5743$    & $0,8224$\\
%$\Lambda$CDM & $0,5487$  & $0,8296$\\
Our model & $0.57$    & $0.82$\\
$\Lambda$CDM & $0.55$  & $0.83$\\
\hline
\end{tabular}
\caption{Comparison of $\gamma$ and $g$ in our model\\ and in the standard $\Lambda$CDM at redshift $z\approx1$.}
\label{tbl:g_dan_gamma} % spaces are big no-no withing labels
\end{table}

\subsection{Numerical exploration}

We now turn to numerical explorations to confirm the analytical results discussed in the previous subsection. We express the perturbation equations in a synchronous gauge and then predict matter and CMB power spectra following our model. Finally, we calculate the $\sigma_8$ parameter, i.e., the r.m.s. of density fluctuation at the $8h^{-1}M\!pc$ scale.

\subsubsection{Perturbations in synchronous gauge}
Let us start from the conformal time coordinates $\eta$. We apply gauge transformation to obtain perturbation equations in the synchronous gauge. The relation between gravitational potential $\Phi$ in the conformal Newtonian gauge and the trace part of perturbation metric $h_T$ in the synchronous gauge is $h_T=\frac{\Phi}{6}$ \cite{Ma_1995}. Using this relation, in the synchronous gauge, Equations (\ref{eq:pers. gangguan kerapatan delta longitudinal}) and (\ref{eq:pers. divergensi kecepatan theta longitudinal}) become \cite{Ma_1995}

\small
\begin{equation}
    \frac{d\delta_X}{d\eta} = 3\mathcal{H}\left(w_X-c_{s,X}^2\right)\delta_X -\left(\theta_{k,X}+\frac{1}{2}\frac{d h_T}{d\eta}\right)\left(1+w_X\right),
    \label{pers. gangguan kerapatan delta synchronous}
\end{equation}
\normalsize

\noindent
and

\begin{equation}
    \frac{d\theta_{k,X}}{d\eta} = -\mathcal{H}\left(1-3w_X\right)\theta_{k,X} - \frac{\theta_{k,X}}{1+w_X}\frac{dw_X}{d\eta} + \frac{c_{s,X}^2 k^2}{1+w_X}\delta_X.
    \label{pers. gangguan kecepatan theta synchronous}
\end{equation}

\noindent
Here, $\theta_k$ replaces $\hat{\theta}$ in the previous discussion, and $k$ is the perturbation wavenumber.

\vspace{4.0mm}
\noindent
\emph{\textbf{Cold dark matter (CDM)}}
\vspace{3.0mm}

CDM interacts with other particles or components only through gravity. On the cosmological scale, it might behave as a pressureless fluid. In the synchronous gauge, the CDM acts as a reference for synchronous coordinates such that the peculiar velocity vanishes. Setting $\theta_{k,X} = \theta_{k,c} = 0$ and $w_X=w_c=c_{s,X}^2=c_{s,c}^2=0$, we find that Equations (\ref{pers. gangguan kerapatan delta synchronous}) and (\ref{pers. gangguan kecepatan theta synchronous}) become

\begin{equation}
    \frac{d\delta_c}{d\eta} = - \frac{1}{2}\frac{d h_T}{d\eta},
    \label{eq:pers. gangguan delta CDM}
\end{equation}

\noindent
and

\begin{equation}
    \frac{d\theta_{k,c}}{d\eta} = 0,
    \label{eq:pers. gangguan theta CDM}
\end{equation}

\noindent
where the subscript $c$ refers to cold dark matter.

During the matter-dominated era, freezing domain walls act like CDM, so that $\Omega_{eff} = \Omega_m + \Omega_{dw} = \Omega_b + \Omega_c + \Omega_{dw} \approx 1$ or $ \Omega_c+\Omega_b < 1$, where the subscript $b$ denotes baryon. If the domain walls do not cluster, then $\delta_T = \Omega_c\delta_c + \Omega_b\delta_b$.  Consequently, based on Equation (\ref{eq:potensial dari EFE-1 dan EFE-2 subhorizon}), the gravitational potential $\Phi$ in our model would be smaller than that of $\Lambda$CDM. Since $h_T=\frac{\Phi}{6}$, we expect the change in the value of $\Omega_c$ to affect the value of $h_T$ and $\delta_c$.

\vspace{4.0mm}
\noindent
\emph{\textbf{Baryonic matter}}
\vspace{3.0mm}

After neutrino decoupling, baryons (including electrons, in a looser astronomical term) became nonrelativistic. If not coupled with radiation, they behave like pressureless fluids. Hence, $c_s^2$ and $w$ in Equations (\ref{pers. gangguan kerapatan delta synchronous}) and (\ref{pers. gangguan kecepatan theta synchronous}) would vanish. However, the last term of Equation (\ref{pers. gangguan kecepatan theta synchronous}) cannot be neglected for large $k$. If $k$ is large enough, the multiplication between $c_s^2$ (nearly zero) and $k$ preserves the appearance of the acoustic features of $\delta_b$. 

In addition, unlike CDM, for the large $k$ values that correspond to high redshifts $z$, baryons interact electromagnetically  with photons by sharing energy-momentum through Thomson scattering. For baryons, this scattering affects the magnitude of the electron velocity distribution. Electrons interact with protons via Coulomb forces, and since the two kinds of particles have different masses, they generally move at different velocities. Hence, in the synchronous gauge in which the CDM velocity is zero, the average baryon velocity would be $\vec{v}_b = \vec{v_e} - \vec{v_p}$, or $\theta_{k,b} = \theta_{k,e} - \theta_{k,p}$. This nonzero net velocity would affect the magnitude of momentum transfer between baryons and photons as well as the Thomson scattering rate $\Gamma=a n_e \sigma_T$. Hence, the total momentum conservation in the interactions requires an additional term in Equation (\ref{pers. gangguan kecepatan theta synchronous}) by \cite{Ma_1995}

\begin{equation*}
    \frac{4\bar{\rho}_{\gamma}}{3\bar{\rho}_b}an_e\sigma_T\left(\theta_{k,\gamma}-\theta_{k,b}\right).
\end{equation*}

\noindent
Therefore, Equations (\ref{pers. gangguan kerapatan delta synchronous}) and (\ref{pers. gangguan kecepatan theta synchronous}) become

\begin{equation}
    \frac{d\delta_b}{d\eta} = -\theta_{k,b}- \frac{1}{2}\frac{d h_T}{d\eta},
    \label{eq:pers. gangguan delta baryon}
\end{equation}

\noindent
and

\begin{equation}
    \frac{d\theta_{k,b}}{d\eta} = -\mathcal{H}\theta_{k,b}+c_{s,b}^2k^2\delta_b+\frac{4\bar{\rho}_{\gamma}}{3\bar{\rho}_b}an_e\sigma_T\left(\theta_{k,\gamma}-\theta_{k,b}\right).
    \label{eq:pers. gangguan theta baryon}
\end{equation}

Unlike CDM fluctuations that directly depend on noncanonical domain walls that behave like dark matter, baryon fluctuations are not directly affected by noncanonical domain walls because noncanonical domain walls do not share energy density with baryons. However, the total gravitational potential $\Phi$ of baryons and CDM decrease since $\Omega_b + \Omega_c <1$ and CDM is coupled to domain walls through this potential. Therefore, the fluctuations in the baryon density would eventually also be affected. Although Equations (\ref{eq:pers. gangguan delta baryon}) and (\ref{eq:pers. gangguan theta baryon}) are not directly affected by $\Phi$, changing the total potential $\Phi$ would alter $h_T$ and $\frac{dh_T}{d\eta}$  through the relation $h_T=\frac{\Phi}{6}$, giving rise to a change in $\delta_b$.

\vspace{4.0mm}
\noindent
\emph{\textbf{Photons and neutrinos}}
\vspace{3.0mm}

Noncanonical domain walls affect temperature fluctuations of photons through a mechanism similar to that of baryons, namely, through the perturbation metric $h_T=\frac{\Phi}{6}$. Therefore, no changes or additional terms are needed in the equation of temperature fluctuations. The equations are similar to those of $\Lambda$CDM. The same applies to neutrinos, which we take as massless and come in three flavors.

\subsubsection{Einstein-Boltzmann equations}

The perturbation equations in the form of $\delta$ and $\theta$ in the previous discussion (Equations (\ref{eq:pers. gangguan delta CDM}), (\ref{eq:pers. gangguan theta CDM}), (\ref{eq:pers. gangguan delta baryon}), (\ref{eq:pers. gangguan theta baryon})) are known as Einstein-Boltzmann equations, i.e.,  a system of first-order differential equations involving cosmological variables that evolve with time. Following the integration method of Seljak and Zaldarriaga \cite{Zaldarriaga_1998}, we compactly write the Einstein-Boltzmann equations in a general equation as

\begin{equation*}
    \frac{dY}{d\eta} = f\left( \eta,Y,\beta,\Omega_m,\Omega_b,\Omega_T,h \right),
\end{equation*}

\noindent
where
\begin{center}
\begin{equation*}
\begin{split}
    &\eta = \textrm{conformal time};\\
    &Y = \textrm{a set of variables};\\
    &\beta^2 = k^2+K, \textrm{generalized wavenumber for non-flat}\\
    &~~~~~~~~~\textrm{geometry, and}~K=0~\textrm{for a flat universe};\\
    &\Omega_m = \textrm{matter density parameter};\\
    &\Omega_b = \textrm{baryon density parameter};\\
    &h = \textrm{dimensionless Hubble factor}.
\end{split}
\end{equation*}
\end{center}

The set of involved variables $\left(Y\right)$ consists of variables that are directly related to either background or perturbation dynamics. The perturbation-related variables include either physically meaningful variables such as temperature fluctuation and matter density contrast or physically meaningless variables such as perturbation variables of the synchronous gauge. The variable that designates the dynamics of the background is the scale factor $a$. Thus, expressed in its components, matrix $Y$ reads

\begin{equation*}
    Y = \left[ a~~h_T~~h_{TL}~~\delta_c~~\delta_b~~v_b~~n_{\gamma}~~\Theta^{(0)}~~\Theta_8~~\Delta_{\Theta}^{(8)} \right]^T,
\end{equation*}

\noindent
with

\begin{center}
\begin{equation*}
\begin{split}
    &a = \textrm{scale factor;}\\
    &h_T = \textrm{trace part of metric perturbation;}\\
    &h_{TL} = \textrm{traceless part of metric perturbation;}\\
    &\delta_c = \textrm{CDM fluctuation;}\\
    &\delta_b = \textrm{baryonic matter fluctuation;}\\
    &v_b = \textrm{peculiar velocity of baryonic matter;}\\
    &n_{\gamma} = \textrm{neutrino multipoles} \\
    &~~~=\left[ n^{(1)}~~n^{(2)}~~n^{(3)}~~n^{(4)}~~n^{(5)}~~n^{(6)}~~n^{(7)}~~n^{(8)} \right]^T;\\
    &\Theta^{(0)} = \textrm{radiation monopole;}\\
    &\Theta_8 = \textrm{radiation multipoles} \\ &~~~=\left[\Theta^{(1)}~~\Theta^{(2)}~~\Theta^{(3)}~~\Theta^{(4)}~~\Theta^{(5)}~~\Theta^{(6)}~~\Theta^{(7)}~~\Theta^{(8)} \right]^T;\\
    &\Delta_{\Theta,8} = \textrm{radiation multipoles polarization} \\
    &~~~=\left[ \Delta_{\Theta}^{(0)}~~\Delta_{\Theta}^{(1)}~~\Delta_{\Theta}^{(2)}~~\Delta_{\Theta}^{(3)}~~\Delta_{\Theta}^{(4)}~~\Delta_{\Theta}^{(5)}~~\Delta_{\Theta}^{(6)}~~\Delta_{\Theta}^{(7)}~~\Delta_{\Theta}^{(8)} \right]^T.
\end{split}
\end{equation*}
\end{center}

The set of Einstein-Boltzmann equations consists of more than thirty first-order differential coupled to each other. The complete solutions of these equations can be obtained by using integrators or Boltzmann solvers such as \texttt{CMBFAST} \cite{Seljak_1996} and \texttt{CAMB} \cite{Lewis_2000}. The two Boltzmann solvers work in the synchronous gauge. Although physically the synchronous gauge has an independently meaningless variable, it ensures very stable computing processes.

We employed \texttt{CMBACCUR} \cite{cmbaccur}, a Boltzmann solver freeware code written by D. Vangheluwe in \texttt{MATLAB}. Similar to \texttt{CMBFAST} and \texttt{CAMB}, \texttt{CMBACCUR} is based on the direct line-of-sight integration approach of \cite{Seljak_1996}. The \texttt{CMBACCUR} default program does not apply to alternate cosmological models. Hence, we modified most of the modules for our purpose. Because the default program does not accommodate dynamic dark energy models, we first adjusted the EoS parameter. Then, we also changed some derived equations, namely, the matter density equation (perturbation equations), Friedmann's equations (age of the universe), conformal time - scale factor relation, structure growth factor, and recombination.

\subsubsection{Input parameters, normalizations, corrections and initial conditions}

We take the best-fitting parameters of the $\Lambda$CDM model from \cite{planck_2020} as input parameters for our numerical calculations, as shown in Table \ref{tbl:input_parameter}.

\begin{table}[h!]
\centering
\begin{tabular}{|c|c|c|}
\hline
\textbf{Parameter} & \textbf{Value} & \textbf{Description}\\
\hline
$\Omega_T$ & $1$    & Total density parameter\\
$h$ & $0.67$  & Hubble factor\\
$\Omega_{m,0}$ & $0.31$ & Matter density parameter\\
$\Omega_{b,0}$ & $0.05$ & Baryon density parameter\\
$n_s$ & $0.96$ & Scalar spectral index\\
$\tau_{rei}$ & $0.052$ & Optical depth at reionization\\
\hline
\end{tabular}
\caption{$\Lambda$CDM best-fitting parameters taken from \cite{planck_2020} as input parameters for our numerical calculations.}
\label{tbl:input_parameter} % spaces are big no-no withing labels
\end{table}

Normalization of the predicted spectra is performed on large scales around the time of recombination, in which both large-angle anisotropies and large-scale inhomogeneities are not strongly affected by any microphysics that possibly depends on cosmological models. This scale corresponds to multipole $l=10$ in the power spectrum of observed CMB temperature fluctuations or equivalently $k\approx 0.05$ $hM\!pc^{-1}$ in the matter power spectrum. The physics beyond this scale is related to the joint-fluctuation between gravitational potential $\Phi$ and photon temperature, which leads to the so-called Sach-Wolfe (SW) effect. The spectrum in the SW region is proportional to the density fluctuation at the time of horizon crossing that could be inferred from observations. Frontier calculations of the large-angle CMB power spectrum based on several cosmological models show that the power spectra of those models coincide at $l\approx 10$ \cite{Dodelson:1282338}. Hence, we choose this scale for normalization.

In addition, we made corrections to our calculated power spectra by calibrating the power spectra to the CMB temperature fluctuation data from Planck \cite{planck_2020}. By fixing the density fluctuation at the horizon crossing $\delta_H = 4.47 \times 10^{-5}$ (see, e.g., \cite{Dodelson:1282338}), we obtained amplitude corrections at $l=10$, as shown in Table \ref{tbl:C10}. We assume adiabatic perturbations and spatially Gaussian initial power spectra in our calculations.

\begin{table}[h!]
\centering
\begin{tabular}{ |p{4cm}|p{1.1cm}| }
\hline
\textbf{Cosmological model} & $C_{10}$\\
\hline
$\Lambda$CDM & $1.028$\\
Our model & $0.974$\\
\hline
\end{tabular}
\caption{Amplitude corrections to the power\\ spectra at $l=10$ ($C_{10}$)  for $\delta_H=4.47 \times 10^{-5}$}
\label{tbl:C10} % spaces are big no-no withing labels
\end{table}

\subsubsection{Cosmological power spectra}

\begin{figure}[ht]
\centering
    \includegraphics[width=0.93\columnwidth]{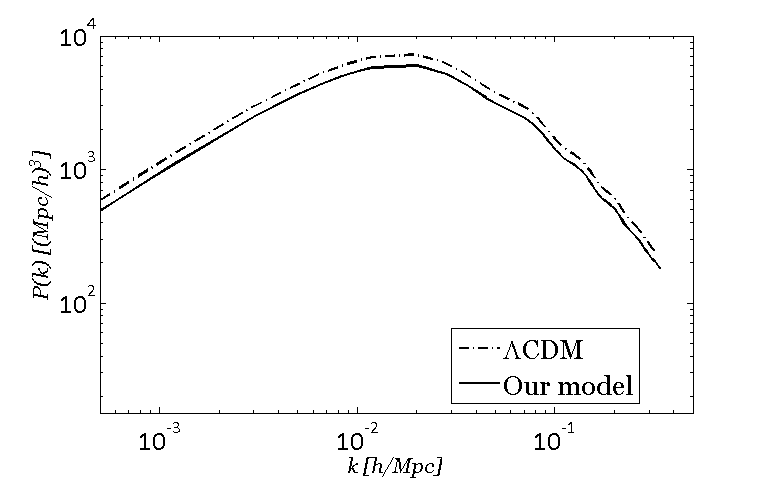}
   \caption{Dark matter power spectra for the noncanonical domain wall model (solid line) and $\Lambda$CDM (dash-dotted line) with input parameters taken from \cite{planck_2020}, i.e., the best-fitting parameters for the $\Lambda$CDM (see Table \ref{tbl:input_parameter})}
    \label{fig:PSDM}
\end{figure}

\begin{figure}[H]
\centering
    \includegraphics[width=0.93\columnwidth]{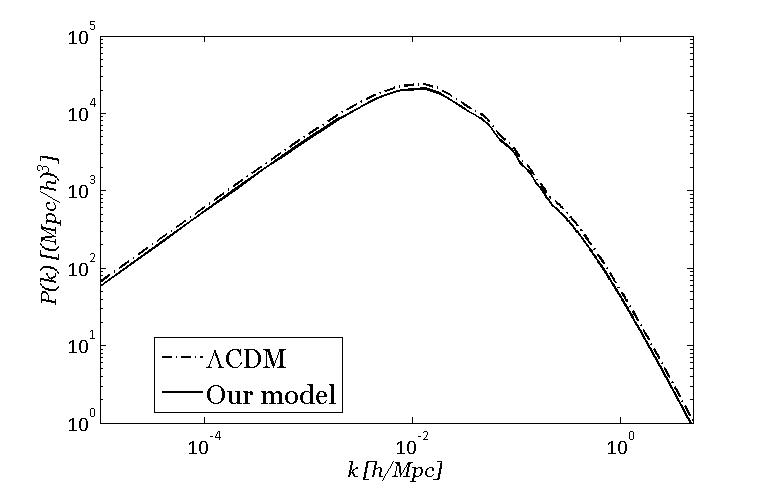}
   \caption{Matter power spectra (baryon and dark matter) for the noncanonical domain wall model (solid line) and $\Lambda$CDM (dash-dotted line) with input parameters taken from \cite{planck_2020}, i.e., the best-fitting parameters for the $\Lambda$CDM (see Table \ref{tbl:input_parameter})}
    \label{fig:PSM}
\end{figure}

Figures \ref{fig:PSDM} and \ref{fig:PSM} show plots of the dark matter and total matter power spectra following noncanonical domain walls (our model) and $\Lambda$CDM, respectively. 

For the set of input parameters shown in Table \ref{tbl:input_parameter}, our model predicts a remarkably similar shape to the $\Lambda$CDM power spectrum, which is slightly lower at all scales. The gap is foreseeable since we know from the previous discussion that nonclustering domain walls take a part of $\Omega_m$, hence suppressing the matter power spectrum. In Figure \ref{fig:PSM}, we linearly extrapolate the spectrum toward smaller scales to reach the part accessible to direct  measurements.

\begin{figure}
\centering
    \includegraphics[width=0.93\columnwidth]{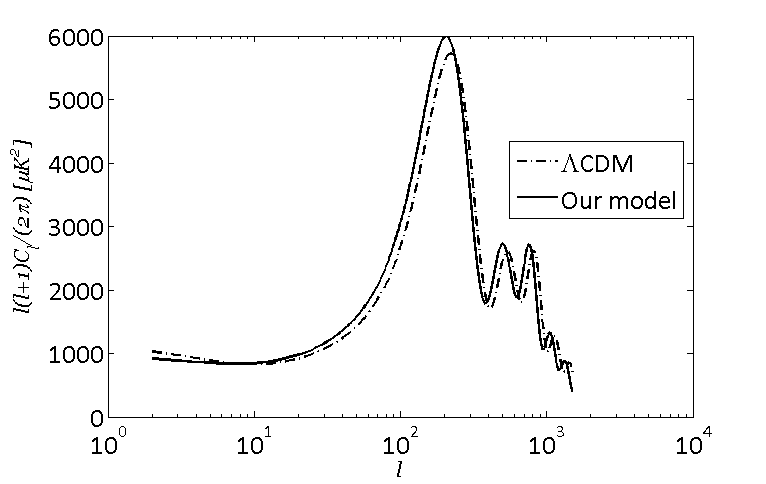}
   \caption{The CMB power spectra ($TT$) for the noncanonical domain wall model (solid line) and $\Lambda$CDM (dash-dotted line) with input parameters taken from \cite{planck_2020}, i.e., the best-fitting parameters for the $\Lambda$CDM (see Table \ref{tbl:input_parameter})}
    \label{fig:PSCMBTT}
\end{figure}

Figure \ref{fig:PSCMBTT} shows a joint plot of the temperature anisotropy power spectra following our model and $\Lambda$CDM. We point out two results from this figure. First, the shape of the power spectrum in our model replicates all features in the $\Lambda$CDM  spectrum. Second, the power spectrum in our model is slightly shifted to smaller angular scales $l$.
 Figure \ref{fig:PSCMBTT} also shows that the anisotropy spectrum is slightly higher on large scales in a noncanonical domain walls universe.

\subsubsection{Inhomogeneity at the scale of  $8~Mpc/h$}
Let us take a closer look at the perturbations at the scale of $8M\!pc/h$. This scale becomes interesting since inhomogeneity on this scale is accessible for direct measurements, independent of cosmological models.
The parameter that describes the inhomogeneity on a scale of $8M\!pc/h$ is
$\sigma_8$.  
To calculate $\sigma_8$ we need an extrapolation since the solution of the Einstein-Boltzmann equations does not reach this scale.
A linear extrapolation of the matter power spectrum is given by \cite{DODELSON2021195}:

\begin{equation}
    P(k,a) = 2\pi^2 \delta_H^2 \frac{k^n}{H_0^{n+3}}T^2(k) \left( \frac{D_1(a)}{D_1(a=1)} \right)^2,
    \label{eq: P(k)}
\end{equation}

\noindent
where $T(k)$ is the matter transfer function on a small scale, and $D_1(a)$ is the growth factor given by

\begin{equation}
    D_1(a) = \frac{5\Omega_m}{2} \frac{H(a)}{H_0} \int_0^a \frac{da'}{\left(a'H(a')/H_0\right)^3}.
    \label{eq: D_1(a)}
\end{equation}

\noindent
The values of $T(k)$ and $D_1(a)$ depend on the choice of the cosmological model.

Using the input parameters in Table \ref{tbl:input_parameter}, we obtain $\sigma_8 \approx 0.76$ in our model, which is smaller than $\sigma_8 \approx 0.83$ in the $\Lambda$CDM model.  This result needs further study to determine its possibility of relieving the $\sigma_8$ tension.

\section{Conclusions and Future Work}

We have investigated domain walls with a noncanonical Lagrangian shown in Equation (\ref{eq:L2}). We found that the EoS of the domain walls depends on their velocity in the observer's frame. For all allowed values of $v$, the noncanonical domain walls might serve as candidates for a unification model of dark matter and dark energy. In early times, domain walls behave like dark matter, whereas in late time, they act as dark energy. We show that the dark matter and dark energy properties of the domain walls depend on their velocity $v$ in the observer's frame.
Domain walls with velocity $v$ larger than the characteristic velocity $v_{ph}$ behave like an ordinary scalar field with $w_{dw} \geq -1$, whereas those with $v<v_{ph}$ in late time possess phantom field properties with $w_{dw}<-1$.

Domain walls with a velocity allowed by special relativity satisfy the stability conditions of classical dynamics and special relativity. However, only freezing domain walls with $v=0$ fulfill all the energy stability conditions of quantum mechanics. Therefore, to avoid instability, domain walls must immediately be frozen after being created; otherwise, they physically never exist.

As dark energy, freezing domain walls generate late-time cosmic acceleration, starting from $z \approx 0.24$, slightly later than the acceleration onset by the cosmological constant, and give the current value of the EoS parameter, $w_{dw,0}=-1.5$. This value is in good agreement with the result of the CMB Planck measurements for dataset \textit{TT}, \textit{TE}, \textit{EE}, low \textit{E}, and lensing, and its combinations \cite{planck_2020}. Our model predicts a slightly slower growth rate of structures compared to that of the $\Lambda$CDM scenario. This is confirmed by the prediction of the dark matter power spectrum following this model, which also shows smaller values at all scales compared to $\Lambda$CDM. These mismatches are related to the dark energy-dark matter unification properties of the domain walls, which effectively counts as dark matter during the matter-dominated era but does not contribute to gravitational collapse because, as solitons, they are not clustered. On small scales, we obtain a smaller value of $\sigma_8$ compared to that in $\Lambda$CDM, which needs further investigation to see its possibility of relieving the $\sigma_8$ tension. Our model  replicates all features in the $\Lambda$CDM CMB spectrum; only the features shift slightly to a smaller angular scale $l$. 

In a following study, we will explore further physical aspects of the domain walls to answer more basic questions concerning dark energy, such as the physical mechanisms behind cosmic acceleration, or whether dark energy is composed of physically perceivable dark energy particles. Beyond the topics to be covered in the following work, large inquiries remain to be addressed regarding what might occur to the unstable domain walls; that is, whether they decay into other entities or they never exist. Even if they could exist, would they be able to exist indefinitely in a negative energy state through some unknown physical mechanism, or would they freeze in some way? If they can exist in a negative energy state, there must be an explanation regarding the cause of the boost at the beginning of time so that their velocity does not vanish. These questions are fundamental questions that are open to the community.

\acknowledgments

This research was funded by the P2MI 2021 project of the Faculty of Mathematics and Natural Sciences, Institut Teknologi Bandung (ITB).
We thank Dr. Premana W. Premadi from Bosscha Observatory - Institut Teknologi Bandung (ITB) and Prof. Dr. Triyanta from Theoretical High Energy Physics Research Group - Institut Teknologi Bandung (ITB) for beneficial advice and discussions.

%\paragraph{Conflict of Interest.} The authors does not have any financial or proprietary interests in any material discussed in this article.

%\paragraph{Funding.} This research was funded by the P2MI 2021 project of the Faculty of Mathematics and Natural Sciences, Institut Teknologi Bandung (ITB).

%\paragraph{Data Availability.} The datasets generated during and/or analysed during the current study are available from the corresponding author on reasonable request.

% The bibliography will probably be heavily edited during typesetting.
% We'll parse it and, using the arxiv number or the journal data, will
% query inspire, trying to verify the data (this will probalby spot
% eventual typos) and retrive the document DOI and eventual errata.
% We however suggest to always provide author, title and journal data:
% in short all the informations that clearly identify a document.

\end{document}